\documentclass [12pt]{article}
\usepackage{doublespace}
\usepackage{epic}
\usepackage{eepic}
\usepackage{hangcaption}
\usepackage{graphics}
\renewcommand{\baselinestretch}{1}

\title{Renormalons Beyond One-Loop}
\author{ Taekoon Lee 
        \\
        \\  
      Department of Physics \\
      Purdue University\\
      West Lafayette, IN 47907
     }

\date{}
\begin{document}
\maketitle
\newcommand{\be}{\begin{equation}}
\newcommand{\ee}{\end{equation}}
\newcommand{\bear}{\begin{eqnarray}}
\newcommand{\eear}{\end{eqnarray}}
\begin{abstract}
Higher order renormalons beyond the chain of one-loop bubbles are
discussed.  A perturbation method  for the infrared renormalon 
residue is found. The large
order behavior of the current-current correlation function due to
the first infrared renormalon is determined in both QED and QCD to
the first three orders.

\end{abstract} 

\def\thepage{PURD-TH-96-09}
\thispagestyle{myheadings}
\newpage
\renewcommand{\baselinestretch}{2}
\pagenumbering{arabic}
\addtocounter{page}{0}

\section{Introduction}
	
	Perturbation theory in field theories is generally plagued
by the rapidly growing coefficients, which cause the series in weak
coupling to be  asymptotic. Classical solutions,
instantons, cause the perturbative coefficients to grow as $n!$ for
large $n$, where  $n$ is the order of perturbation, and so do certain
subsets of Feynman diagrams, renormalons\footnote{ 
Renormalons  also denote  the singularities in the Borel plane.}.
Some of the properties  of the renormalons are discussed here. 

A chain of the one-loop bubble diagrams  in a photon
propagator in massless QED is an example of the renormalons (Fig.1).
An exchange of the one-loop Gell-Mann--Low (GL) effective charge
 gives a contribution of $ n!$  for an ultraviolet
renormalon, in which the momentum flowing in the propagator is
large  compared to the renormalization scale ($k^{2}/\mu^{2} \sim
\exp{n}$), and $ (-1)^{n} n!$ for
an infrared (IR) renormalon, in which a soft momentum ($ k^{2}/\mu^{2}
\sim\exp{(-n/2)}$)
flows in the  propagator.

The actual form of the large order behavior due to an infrared renormalon
is generally given by
\begin{equation}
	K\, n!\, n^{\nu}\, b_{0}^{-n}\, ( 1 + O(1/n)),
\label{1} \end{equation}
where $\nu$ and $b_{0}$ are renormalon-specific, known constants.
The coefficient $K$ is an all-order quantity \cite{grunberg1}.
 It depends not only
on the one-loop renormalon  mentioned above but also on an infinite
set of higher order renormalons, and so determining it is nontrivial.

  However, it should be emphasized that $K$ is calculable, at least 
perturbatively. For example, if we had calculated the series to a very
high order, then eq. (\ref{1}) implies that we could  extract the
coefficient to an accuracy of $O(1/n)$.
Therefore there must
be a convergent sequence $K_{N}$  for $K$, with $K$ being its limit,
associated with the perturbation of the amplitude in consideration.
The main purpose of this paper is to present such a 
sequence for the first IR renormalon  in the Borel plane.

 The precise calculation of 
the large order behavior is important, besides its theoretical interest,
 because it could play an essential role
  in an effort to  reconstruct the 
true amplitudes from the  perturbation theory.
 The large order behavior
 due to the IR renormalons
 in nonabelian gauge
theory  arises from  the imaginary part of the
 nonperturbative effects, vacuum condensations, 
 and so a precise calculation of the
large order behavior gives  detailed information on the
imaginary part of the nonperturbative amplitudes, which could be 
 essential in understanding the full amplitude. For a recent
consideration in this direction one may refer to \cite{grunberg2}.

This paper is organized as follows.
In sec. 2, we discuss in QED the higher order renormalons  
beyond the chain of 
one-loop bubbles, and show in detail  how the large order
 behavior gets contribution
from the higher order renormalons. In sec. 3--5, a systematic
method of summing those higher order renormalons is discussed, and the 
renormalon residue to the first three orders is given. In sec. 6, we
discuss the calculation of  the large order behavior  in QCD using
the analytic property in Borel plane, and give the large order behavior
to the first three orders. In sec. 7, the scheme dependence of the 
large order behavior is discussed.

\begin{figure}[ht]
\leavevmode
\begin{center}
\rotatebox{0}{
\resizebox{4.5cm}{5cm}{%
\includegraphics* [2cm,4cm][20cm,25cm]{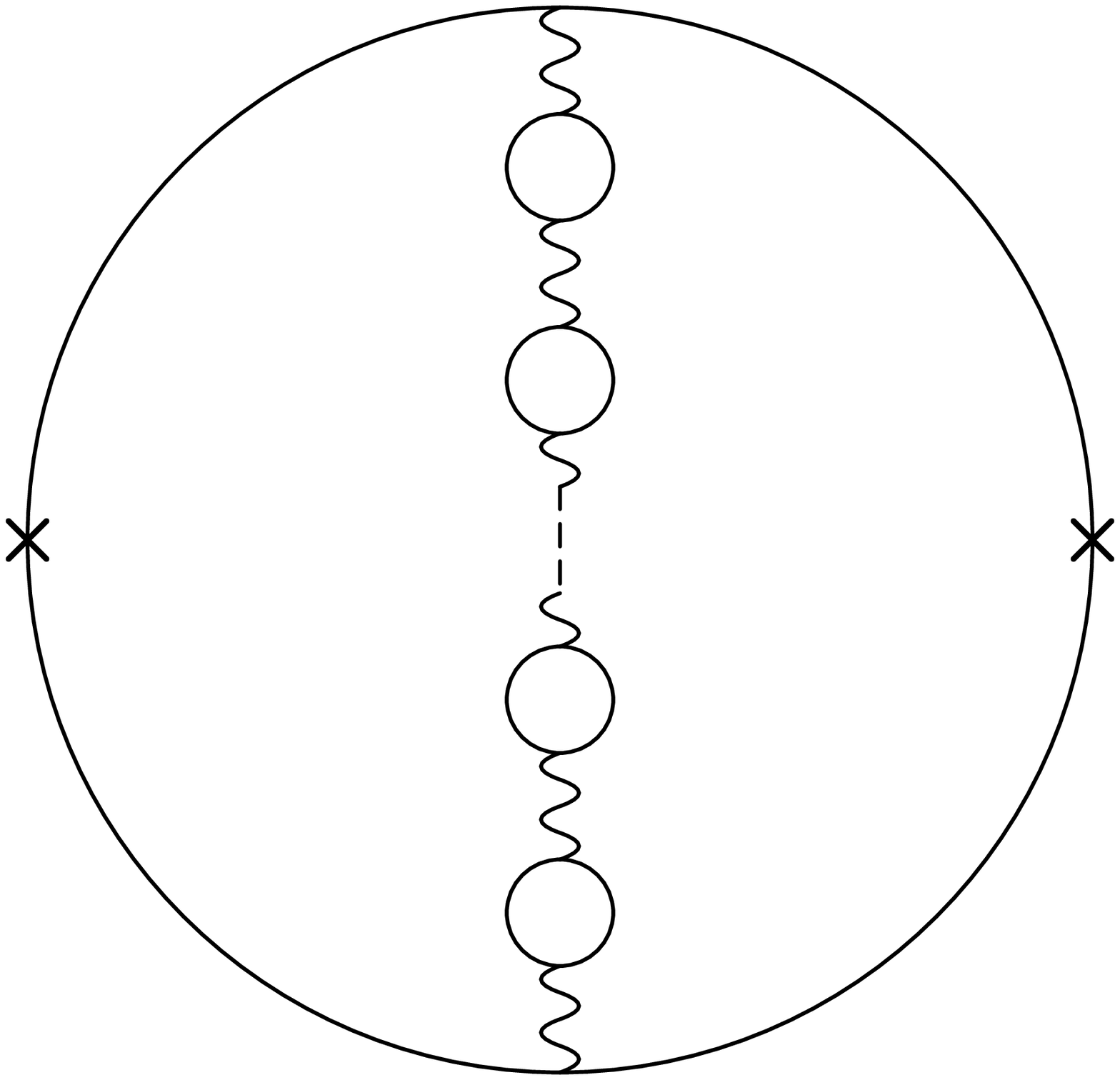}} } 
\end{center} 
\isucaption{One-loop renormalon.}
\end{figure}

\section{Higher order renormalons}

We first review how a chain of the one-loop bubble diagrams gives
rise to  factorial growing coefficients, and then show that the large
order behavior of perturbation theory is an all order property by
giving an estimate  of the higher order renormalons.

The Green's function we consider is the electromagnetic current
correlation function in QED in the Euclidean regime
\begin{equation}
i \int e^{i q x} < j_{\mu}(x) j_{\nu}(0) > d^{4} x =
(q_{\mu}q_{\nu}-q^{2} g_{\mu\nu}) \frac{\Pi(\alpha(\mu),
\frac{\mu^{2}}{Q^{2}})}{\alpha(\mu)},
\label{2} \end{equation}
where $Q^{2} = - q^{2} > 0$, and
\begin{equation} 
 j^{\mu}(x)=\bar{\psi} \gamma^{\mu} \psi(x).
\label{3} \end{equation} 
For large order behavior, it is more convenient to consider the 
renormalization scheme and scale invariant quantity $D$ defined by 
\begin{eqnarray}
D ( \alpha(\mu), \mu^{2}/Q^{2} ) & = &\left. Q^{2} \frac{\partial}{
\partial Q^{2}} \left[  \frac{\Pi(\alpha(\mu),\frac{\mu^{2}}{
Q^{2}})}{\alpha(\mu)}\right] - Q^{2} \frac{\partial}{
\partial Q^{2}} \left[  \frac{\Pi(\alpha(\mu),\frac{\mu^{2}}{
Q^{2}})}{\alpha(\mu)}\right] \right|_{\alpha(\mu)=0} \nonumber \\
& = & \sum_{n=0}^{\infty}a_{n}(\mu^{2}/Q^{2})[\alpha(\mu)]^{n+1}.
\label{4} \end{eqnarray}

A single exchange of the GL effective charge gives rise
to the first IR renormalon singularity in the Borel plane. 
It is also generally assumed that the leading  residue of the
first IR singularity can be  completely determined by a 
single exchange of the GL effective charge, which
implies equivalently that the large order behavior due to the
 IR renormalon can be
determined by a single exchange of the GL effective
charge ( Fig.1 ). For the other IR renormalon in the Borel plane,
it  is similarly  believed that their residues
 can be saturated  by the multiple exchanges of
the GL effective charges. The analysis for IR renormalon
 in nonabelian gauge theory
 using operator product expansion supports this assumption 
\cite{parisi, mueller1}.
With this  assumption for the  large order behavior  
 for the first IR renormalon,
 $D$ may be written  as 
\begin{equation}
D(\alpha(\mu),\mu^{2}/Q^{2}) = \int_{0}\, f( k^{2} )\,
\mathbf{a} (k^{2})\, d\, k^{2}
\label{5} \end{equation}
where $ \mathbf{a} ( k^{2} )$ denotes  the GL effective charge and
\begin{equation}
f (k^{2})  =   \frac{ -e_{f}^{2} k^{2}}{
8 \pi^{3} Q^{4}} \hspace{.5in}  \mbox{ for $k^{2} \rightarrow 0$}
\label{6} \end{equation}
with $e_{f}$ denoting  the  charge of the fermion $\psi$.
This infrared limit of $f(k^{2})$ can be easily read off from
the coefficient of $F_{\mu\nu}^{2}$ term in the operator product
expansion of the current product in eq. (\ref{2}).

To see the $n!$ growth of the perturbative coefficient from the
chain of the one-loop bubbles, we may substitute $\mathbf{a}(k^{2})$ in
eq. (\ref{5}) with its one-loop form
\begin{eqnarray}
\mathbf{a}(k^{2}) & = & \frac{\alpha(\mu)}{1 - \beta_{0} \alpha(\mu)
 \ln \left( \frac{k^{2}}{ \mu^{2}}\right)} \nonumber \\
         &=& \sum_{0}^{\infty} \left[ 
\beta_{0} \ln \left( \frac{k^{2}}{ \mu^{2}}\right) \right]^{n} 
\alpha(\mu)^{n+1},
\label{9} \end{eqnarray}
where $\beta_{0}$ is the first coefficient of the $\beta$ function,
to obtain 
\begin{eqnarray}
a_{n}  &=& -\frac{ e_{f}^{2} \mu^{4}
 \beta_{0}^{n}}{
8 \pi^{3} Q^{4}} \int_{0} \, t \ln (t)^{n} dt \nonumber \\
&=& -\frac{ e_{f}^{2} \mu^{4}}{16 \pi^{3}Q^{4}} \left( -\frac{\beta_{0}}{2}\right)
^{n} n ! ( 1 + O(1/n)),
\label{10} \end{eqnarray} 
where
\begin{equation}
t = \frac{k^{2}}{\mu^{2}}.
\label{11} \end{equation} 
For large $n$ the leading  contribution to the integral comes 
from the kinetic
region $ k^{2} \sim \mu^{2} \exp ( -n/2 )$, and thus the leading 
large order behavior  is independent of the upper
bound of the integral.

Let us now consider the effect of higher order renormalons on
the large order behavior of perturbation. First we  introduce
some definitions. In the following  the vacuum polarization diagrams 
are assumed to include  two
external photon propagators. 
An irreducible renormalon is defined  by replacing
all photon propagators in an irreducible vacuum polarization diagram
with chains of the one-loop bubbles. Similarly, reducible renormalons are
defined by replacing all photon lines in reducible vacuum 
polarization diagrams with chains of the  one-loop bubbles. Thus for
every vacuum polarization diagram there are corresponding 
renormalons. 

\begin{figure}[ht]
\leavevmode
\begin{center}
\rotatebox{0}{
\resizebox{8cm}{10cm}{%
\includegraphics* [2cm,4cm][20cm,25cm]{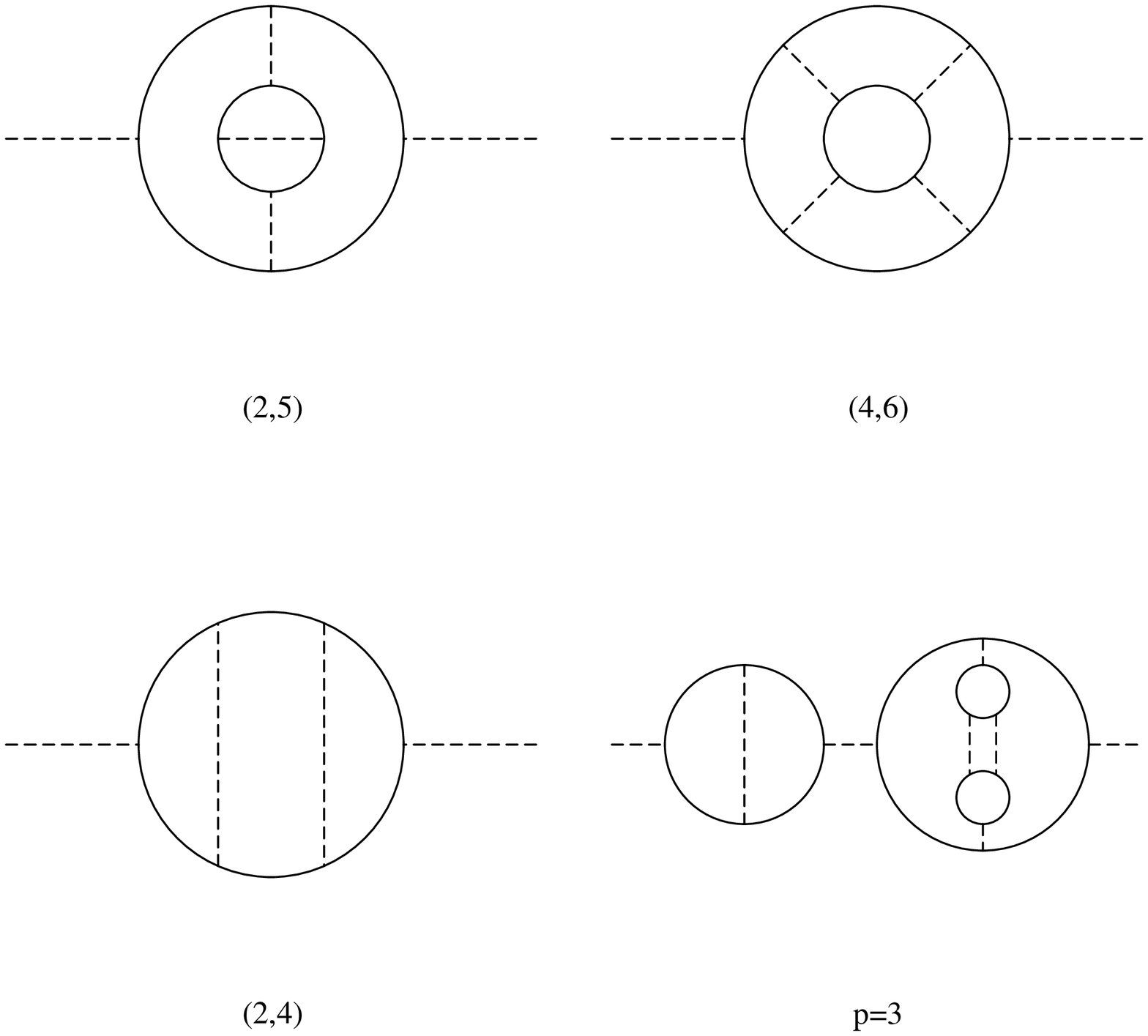}} } 
\end{center}
 \isucaption{Examples of higher order 
 renormalons. Dashed lines denote
chains of one-loop bubbles, and $(p,q)$ denote the
order and number of the reduced photon propagators respectively.}
\end{figure}

We assign an order $p$, and the number of  reduced photon propagators,
$q$, to each irreducible vacuum polarization diagram by
\begin{eqnarray}
p &=& n_{A} -n_{L}-1 \nonumber \\
q &=& n_{A}-n_{1},
\label{12} \end{eqnarray}
where $n_{A}$ is the number of photon propagators and $ n_{L}, n_{1}$
denote the number of irreducible vacuum polarization subdiagrams
and the number of the one-loop bubbles 
respectively. A reduced photon propagator is simply a chain of
an unspecified number of one-loop bubbles. The same $p$ and $q$ of
an irreducible vacuum polarization diagram are defined as the
order and the number of the reduced photon lines of the corresponding
renormalon.  For example, the order and the number of reduced photon
 lines of
the one loop renormalon in Fig.1 is $p=0, q=1$. 
Other higher order
renormalons   may be similarly characterized by the  $p$ and $q$.
For reducible renormalons,  the 
highest order of the irreducible subrenormalons of a reducible
renormalon is defined as the order of the reducible renormalon.
Some examples of the higher order renormalons are given in Fig.2.

\begin{figure}[ht]
\leavevmode
\begin{center}
\rotatebox{0}{
\resizebox{8cm}{10cm}{%
\includegraphics* [2cm,4cm][20cm,25cm]{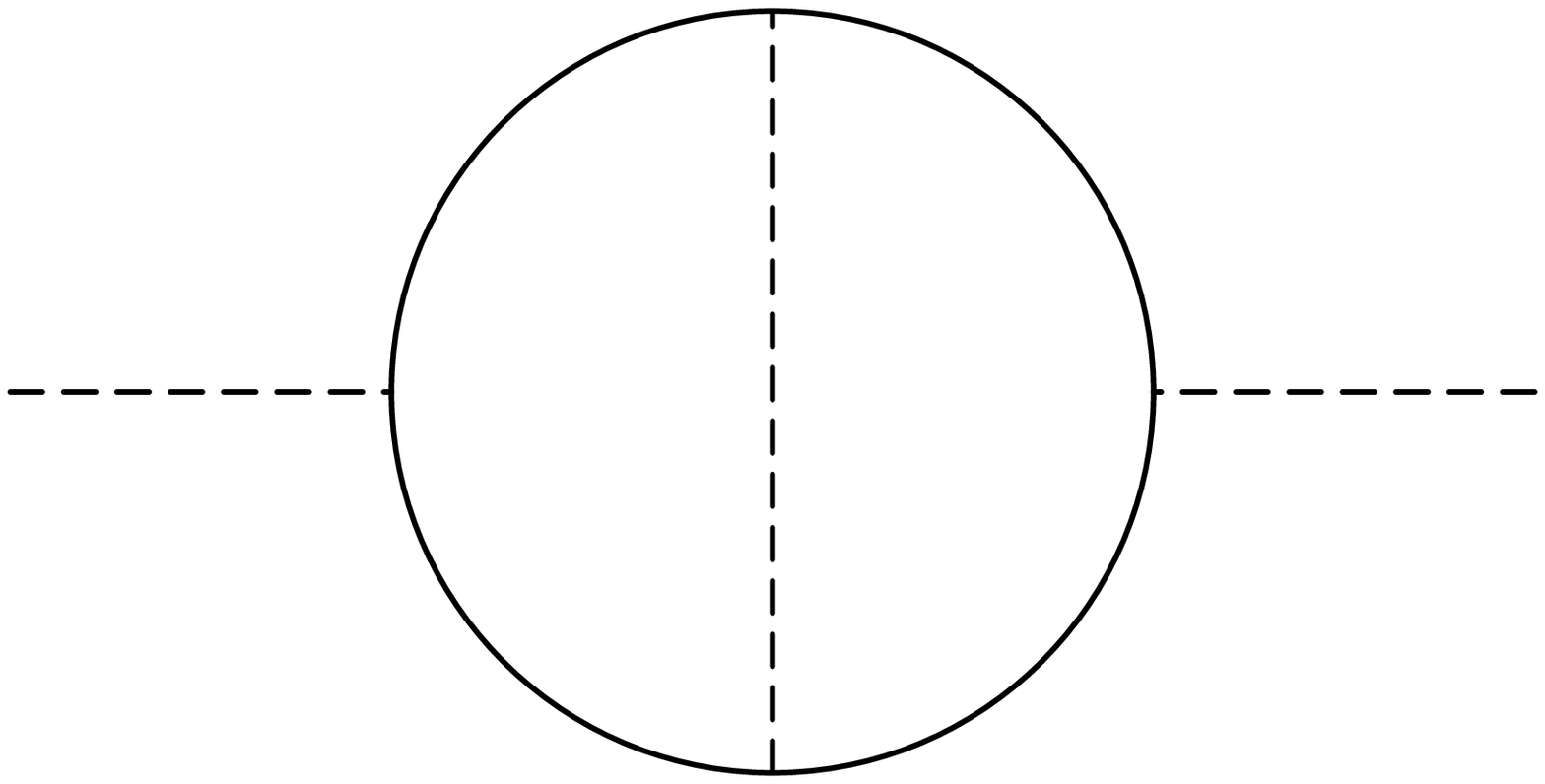}} } 
\end{center}
\isucaption{Order one renormalon. Dashed lines denote chains of
the one-loop bubbles.}
\end{figure}

Before we discuss the effect on large order behavior of 
higher order renormalons in general, let us take
some specific examples of low order renormalons, and see their 
contribution to the large order behavior. This exercise  is very
instructive and gives an insight to more complex renormalons.

It was first noticed by Grunberg \cite{grunberg1}, and
 diagrammatically by Mueller \cite{mueller2}, 
that the large order behavior of perturbation  is an all order
property ( see also \cite{martin}).
 The following argument is motivated by Mueller's observation.
Let us consider the  order one renormalon in Fig. 3. The 
coefficient $\widetilde{\mathbf{a}}_{n}(t)$
 of $ \alpha(\mu)^{n+1}$ due to  this renormalon in the 
perturbation of the GL effective charge in $\alpha(\mu)$ is given by
\begin{equation}
\widetilde{\mathbf{a}}_{n}(t) = \sum_{r=1}^{n-1} (-\Pi_{1}(t))^
{n -r-1}\left[- \Pi_{r +1}^{\left(r-1\right)} (t)\right] ( n-r)
\label{13} \end{equation}
where $\Pi_{r}$ is the $r$-loop vacuum polarization function
\begin{equation}
\Pi(\alpha(\mu),t)= \sum_{r=1}^{\infty} \Pi_{r}(t) \alpha(\mu)^{r}.
\label{14} \end{equation}
The powers of $\Pi_{1}$ in eq. (\ref{13})
 obviously come from the one loop bubbles
in the external reduced photon propagators, and the factor 
$ (n -r) $ accounts for the $ (n - r)$ possible location of 
the fermion loop with  the internal reduced photon line.
$\Pi_{r+1}^{\left(m\right)}$ denotes the terms
proportional to $\beta_{0}^{m}$  in $\Pi_{r+1}$.

 The general form of
$\Pi_{r+1}^{\left(m\right)}$ can be deduced by considering the
following renormalization group equation for $ \Pi ( \alpha(\mu),t)$
\begin{equation}
\mu^{2} \frac{d}{ d \mu^{2}} \left[ \frac{1}{\alpha(\mu)} +
  \frac{1}{\alpha(\mu)} \Pi( \alpha(\mu), t) \right]=0,
\label{15} \end{equation}
which comes  from the renormalization scale invariance of the 
GL effective charge $\mathbf{a}(k^{2})$
\begin{equation}
\mathbf{a}(\alpha(\mu),t) = \frac{\alpha(\mu)}{1 + 
\Pi(\alpha(\mu),t)}.
\label{16} \end{equation}
Putting the perturbative form of $\Pi$ in eq. (\ref{14}) into eq. (\ref{15}) we get
the recursion equation 
\begin{equation}
t \frac{d}{d t}\Pi_{n + 1} = - \beta_{n} + \sum_{m=2}^{n} ( m-1)
\beta_{n-m} \Pi_{m}, \hspace{.5in} \mbox{for } n = 2,3, \cdots
\label{17} \end{equation}
with
\bear
\Pi_{1}&=& -\beta_{0} \ln t + p_{1} \nonumber \\
\Pi_{2}&=& -\beta_{1} \ln t + p_{2}.
\label{18} \eear
Here $\beta_{m}$  are the coefficients of the $\beta$ function
defined by
\be
\beta(\alpha(\mu)) = \mu^{2} \frac{d \alpha(\mu)}{ d \mu^{2}} =
\sum_{m=0}^{\infty} \beta_{m} \alpha(\mu)^{m+2}
\ee 
and $p_{i}$ are constants.

Solving the recursion equation  we have
\begin{equation}
\Pi_{r+1}^{\left(r-1\right)}= ( \beta_{0} \ln(t))^{r-1} \left[ 
\frac{-\beta_{1}}{r} \ln(t) + p_{2}\right].
\label{19} \end{equation}
Substitution of  this into eq. (\ref{13}) gives
\begin{equation}
\widetilde{\mathbf{a}}_{n}( t) = \sum_{r=1}^{n-1} 
( \beta_{0} \ln(t))^{n-2} 
 \left[ \frac{\beta_{1}}{r} \ln(t) - p_{2}\right] (n-r). 
\label{20} \end{equation}
Here we kept only the $\ln (t)$ term in $\Pi_{1}(t)$ for simplicity, and
the effect of the constant term in $\Pi_{1}(t)$ will
 be discussed shortly.
Since a factor of  $ (\ln t)^{n}$ in the integrand in eq. (\ref{5})
would  give rise to 
\begin{equation}
\frac{ (-1)^{n} n!}{2^{n+1}}
\label{21} \end{equation}
$\widetilde{\mathbf{a}}_{n} ( t) $ gives the 
following large order behavior
\begin{equation}
a_{n}(\mu^{2}/Q^{2})= -\frac{ e_{f}^{2} \mu^{4}}{16 \pi^{3} Q^{4}}
 \left( - \frac{\beta_{0}}{2}\right)^{n} n! \left[
\frac{- 2 \beta_{1}}{\beta_{0}^{2}} ( \ln n + \gamma_{E} -1)
 -2 \frac{p_{2}}{\beta_{0}^{2}}
\right] \left( 1 + O \left(1/n\right)\right)
\label{22} \end{equation}
which is comparable to the one-loop renormalon contribution. Here
$\gamma_{E}$ is the Euler constant.

 Going back to eq. (\ref{13}), expanding
 the factor 
\be
(-\Pi_{1})^{n-r-1}=(\beta_{0} \ln t - p_{1})^{n-r-1}
\label{23} \ee
it is easy to see that every term in the expansion proportional to
\be
(p_{1})^{i} \hspace{.5in}\mbox{for\,\,\, $ i \ll n$}
\label{24} \ee
also contributes to the large order behavior. Thus the the
inclusion of the constant term in $\Pi_{1}(t)$ 
would modify eq. (\ref{22}) into a form
\be
a_{n} \sim  \left(-\frac{\beta_{0}}{2}\right)^{n} n!\, \left[\,
\frac{\beta_{1}}{\beta_{0}^{2}}\,\left[\, (\ln n )\, h(p_{1})  +
      h'(p_{1})\,\right]   + \frac{p_{2}}{\beta_{0}^{2}} \,
 h''(p_{1})\,\right]
 \left( 1 + O \left(1/n\right)\right)
\label{25} \ee
where
\be
h(p_{1})= h_{0}+ h_{1} \left(\frac{p_{1}}{\beta_{0}}\right) +
 h_{2}    \left(\frac{p_{1}}{\beta_{0}}\right)^{2}    + \cdots. 
\label{26} \ee 
Here $h_{i}$ are  calculable constants. Note that this series
runs to an infinite order in the limit $n \rightarrow \infty$.
  $h'(p_{1}), h''(p_{1})$ are
similarly defined in a series form.
\begin{figure}[ht]
\leavevmode
\begin{center}
\rotatebox{0}{
\resizebox{8cm}{5cm}{%
\includegraphics* [2cm,6cm][25cm,25cm]{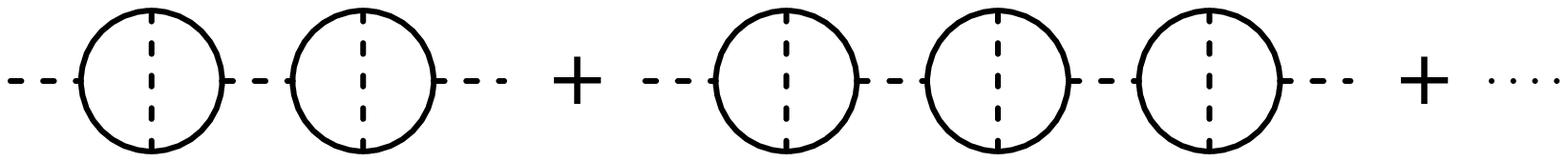}} } 
\end{center}
\isucaption{Chains of the order one renormalons.}
\end{figure}

Using a similar method it is now easy to estimate the large order
 behavior from other renormalons. For example,
 it is straightforward to check that
the chains of the order one irreducible renormalon in Fig.4 along
with the one-loop renormalon contribution in eq. (\ref{10})
 exponentiate the $\ln (n)$ term in eq. (\ref{22}) to
give the $n^{\nu}$ factor in the large order behavior with
\begin{equation}
\nu =\frac{- 2 \beta_{1}}{\beta_{0}^{2}},
\label{27} \end{equation}
which agrees with the well-known
 result for the first IR renormalon \cite{mueller1}.

Now to discuss the higher order  renormalons in general,
  consider
an irreducible  renormalon of order $p$ and reduced photon propagator
 number $q$.
The large order behavior due to this renormalon in the infrared
regime  is given by 
\be
a_{n}\sim \int_{0} \, t\, \sum_{r=q-2}^{n-1}
(\beta_{0} \ln t )^{n-r-1} [\Pi_{r+1}^{
\left(r-q+2\right)}(t)] (n-r) \,d\,t.
\label{28} \ee
Here we   picked up only the $\ln (t)$ term in $\Pi_{1}(t)$ as before.
The general solution of the recursion equation in eq. (\ref{17}) may be organized 
in terms of $p$ and $q$  as
\be
\Pi_{r+1}= \sum_{q'=1}^{r} \beta_{0}^{r-q'}\, \sum_{p=1}^{q'} \sum_{k=0}^{1}
(\ln t)^{ r-p-k+1} \sum_{\{m_{i},\overline{m}_{i}\}} C_{\left\{m_{i},
\overline{m}_{i}\right\}}^{rpqk} \prod_{i=2}^{p+1} \beta_{i-1}^{m_{i}}
p_{i}^{\overline{m}_{i}} 
\label{29} \ee
where 
\be
q'=q -2
\ee
and $m_{i},\overline{m}_{i}$ are nonnegative integers that 
satisfy
\bear
& \sum_{ i\ge 2}
 m_{i}+\overline{m}_{i} = q'-p+1 \nonumber  \\
& \sum_{i\geq 2} \overline{m}_{i} =k \nonumber \\
& \sum_{i\geq 2}( m_{i}+\overline{m}_{i})\cdot( i-1)=p.
\label{30} \eear
Here $p_{i}$ is  the constant term of $\Pi_{i}(t)$.
The $\Pi_{r+1}^{\left(r-q+2\right)}(t)$ of order $p$ renormalon 
is then given by
\be
\Pi_{r+1}^{\left(r-q+2\right)}(t)=\beta_{0}^{r-q +2}  \sum_{k=0}^{1}\,
(\ln t)^{ r-p-k+1} \sum_{\{m_{i},\overline{m}_{i}\}} C_{\left\{m_{i},
\overline{m}_{i}\right\}}^{rpqk} \prod_{i=2}^{p+1} \beta_{i-1}^{m_{i}}
p_{i}^{\overline{m}_{i}}.
\label{31} \ee
By solving   the recursion equation in eq. (\ref{17}) explicitly for the 
several
low order  diagrams, and considering the form of the large order 
behavior in eq. (\ref{1}), it is not difficult to convince oneself that
asymptotically
\be
C_{\left\{m_{i},\overline{m}_{i}\right\}}^{rpqk}
 \sim r^{ p+k-2} g(\ln r), 
\hspace{.4in} \mbox{for $ r \rightarrow \infty$}.
\label{32} \ee
Here the function $g(x)$ is a polynomial of degree at most  $m_{2}-1$ for
$m_{2} \ge 2$, 
and causes the $n^{\nu}$ term in the large order behavior.

Substituting eq. (\ref{31}) into eq. (\ref{28}) and ignoring the
 logarithmic dependence
in $g$, we get
\be
a_{n} \sim \frac{n! \left(- \frac{\beta_{0}}{2}\right)^{n}}{\left(
- \beta_{0}\right)^{q-1}} \sum_{k=0}^{1} J(p,k)
\sum_{\left\{m_{i},\overline{m}_{i}\right\}}\prod_{i=2}^{p+1}
\beta_{i-1}^{m_{i}}p_{i}^{\overline{m}_{i}},
\label{33} \ee
where $J(p,k)$ is a function of $p$ and $k$.
This shows that the irreducible renormalons of all order contribute to
the large order behavior. Further note that the large order
 behavior due to
an irreducible renormalon of order $p$ involves only the coefficients
of the vacuum polarization function and the $\beta$ function to
$(p+1)$-loop order. 

These are also true for the reducible renormalons.
 As we chain more irreducible
subrenormalons into a reducible renormalon, more powers of
$ \ln t$ in $\widetilde{\mathbf{a}}_{n}(t)$ due to this renormalon 
 are being lost,  resulting in suppressed
integral in $t$,
 but this suppression is exactly compensated by the larger
combinatoric factor caused by the more possible locations in putting the
 the
subrenormalons, giving  a large order behavior comparable
to those from the irreducible renormalons. Therefore,
all reducible renormalons also contribute to the leading
 large order behavior.
Also since $ \widetilde{\mathbf{a}}_{n}(t)$ for a reducible renormalon
 with $m$ irreducible subrenormalons
is proportional to 
\be
\Pi_{r_{1}+1}^{\left(r_{1}-q_{1}+2\right)}\Pi_{r_{2}+1}^{
\left(r_{2}-q_{2}+2\right)} \cdots
\Pi_{r_{m}+1}^{\left(r_{m}-q_{m}+2\right)},
\label{34} \ee
with  each $\Pi_{r_{i}+1}^{\left(r_{i}-q_{i}+2\right)} $ coming from the
subrenormalons, it is obvious that the large order behavior from
this reducible renormalon depends on the vacuum polarization 
function and the $ \beta$ function to $(p+1)$-loop order.

The inclusion of the constant term in $\Pi_{1}(t)$ in eq. (\ref{28})
 would modify the large order behavior in 
eq. (\ref{33}) in a similar fashion as in the example of the  order one
irreducible renormalon in eq. (\ref{25}). With $p_{1}$ included, 
each term in eq. (\ref{33}) will be multiplied by a
series in $p_{1}$ in the form of eq. (\ref{26}), with the coefficients
 $h_{i}$  now depending on each  particular term.

Though this  analysis of diagrams  is very helpful in
understanding  general higher order renormalons qualitatively,
it  seems  difficult, or at least inconvenient, 
to systematically calculate the higher order renormalons using this
technique.
We need a  more straightforward approach
for systematic evaluation of the higher order renormalons. In
the following sections such  an approach is discussed.

\section{ Borel transform of GL effective charge}

The problem of determining the renormalon residues of
 $D(\alpha(\mu),\mu^{2}/Q^{2})$  
eventually reduces
to finding the Borel transform of the GL effective charge.
We give in this section a perturbation method for 
the Borel transform of $\mathbf{a}(k^{2})$.
We are going to introduce a scheme and scale independent coupling, and 
then write the GL effective charge in terms of the coupling
in a form that is particularly convenient for Borel 
transform.    

The Borel transform of $D(\alpha(\mu),\mu^{2}/Q^{2})$ is defined by
\be
D(\alpha(\mu),\mu^{2}/Q^{2})=\int_{0}^{\infty} \exp \left(-\frac{b}{
\alpha(\mu)}\right)
\widetilde{D}(b)\,d\,b.
\label{35} \ee
With the perturbative series of $D(\alpha(\mu),\mu^{2}/Q^{2})$ in eq. (\ref{4}),
\be
 \widetilde{D}(b)=\sum_{n=0}^{\infty} \frac{a_{n}}{n!} b^{n}.
\label{36} \ee
The large order behavior of the form in eq. (\ref{1}) causes a
singularity (renormalon)
  in $\widetilde{D}(b)$, and conversely the singularity
in $ \widetilde{D}(b)$ determines the large order behavior. Thus 
by studying $\widetilde{D}(b)$ near the renormalon
  singularities we can
determine the large oder behavior of $D(\alpha(\mu),\mu^{2}/Q^{2})$.

Let us first consider the renormalization group equation for 
the GL effective charge $\mathbf{a}(k^{2})$,
\be
\left( \mu^{2} \frac{\partial}{\partial \mu^{2}} +
\beta(\alpha)\frac{\partial}{\partial\alpha} \right) \mathbf{a}(
\alpha(\mu),t)=0.
\label{37} \ee
Solving the equation we may write the GL effective charge as
\be
 \mathbf{a}(k^{2})= \frac{1}{\frac{1}{A(k^{2})} + C(\mathbf{a}(k^{2}))}
\label{38} \ee
where $ C(\mathbf{a})$ is a scheme independent function.
The effective coupling $A$ is defined by
\be
A(\alpha(\mu),t)=\frac{1}{ - \beta_{0} \left( \ln t +
 \int^{\alpha(\mu)} \frac{1}{ \beta(\alpha)}\,d\,\alpha -\frac{
p_{1}}{\beta_{0}}\right)},
\label{39} \ee
where $ p_{1}$ is defined in eq. (\ref{18}) and the integral of
 $1/\beta(\alpha)$
is defined in the perturbative form
\be
\int^{\alpha(\mu)} \frac{1}{\beta(\alpha)} \,d\,\alpha = -\frac{1}{\beta_{0}
\alpha(\mu)}  -\frac{\beta_{1}}{\beta_{0}^{2}}\ln \alpha(\mu) + 
\cdots.
\label{40} \ee
Since 
\be
A(k^{2})= \alpha(k^{2}) \left[1 - \frac{\beta_{1}}{\beta_{0}}
\alpha(k^{2}) \ln \alpha(k^{2}) +
\cdots\right],
\label{41} \ee
$A$ can be as good an expansion parameter as $\alpha$.
$A(k^{2})$ is also independent of renormalization scheme. The scheme
independence of $A(k^{2})$   can be easily understood by considering
 the difference of the integral in eq. (\ref{39}) between two schemes.
 Since the difference is $\mu$ independent,  it cannot depend 
on $\alpha(\mu)$ and
thus must be a constant. 
It is in fact  given by
\be
\int^{\alpha'(\mu)} \frac{ d\, x}{\beta'(x)} -
\int^{\alpha(\mu)} \frac{ d\, x}{\beta(x)} = \gamma_{1}
\label{42} \ee
where $\gamma_{1}$ is the first coefficient of the relation between
the couplings of the two schemes
\be
\alpha'(\mu)=\alpha(\mu) \left[ 1 + \gamma_{1} \alpha(\mu) + \gamma_{2} \alpha(\mu)^{2} + \cdots \right].
\label{43} \ee
The difference in eq. (\ref{42}) is then canceled by the scheme dependence of
$p_{1}$
\be
p_{1}'=p_{1}+\gamma_{1},
\label{44} \ee
leading to  the scheme independence of $A(k^{2})$. Note that 
a similar effective charge
 was considered in relation to the renormalization
scheme invariant perturbation by Maximov and Vovk \cite{vovk}.

Now $C(\mathbf{a}(k^{2}))$ in eq. (\ref{38}) may be expanded 
in an asymptotic series  as
\be
C( \mathbf{a}(k^{2}) ) = c_{1} \ln ( \mathbf{a}(k^{2}) ) + \sum_{2}^{\infty} c_{i}
\,[ \mathbf{a}(k^{2})]^{i-1}.
\label{45} \ee
The coefficients $ c_{i}$ can be determined by solving perturbatively
 $ \mathbf{a}(k^{2}) $ 
in eq. (\ref{38})  with the help of eqs. (\ref{39}), (\ref{40}) in terms of 
$ \alpha(\mu)$, and comparing it with the 
perturbative expansion of the GL effective charge in $\alpha(\mu)$
using eq. (\ref{16}). Then it is not difficult to see that
$c_{i}$ are given by the renormalization scheme invariant combinations
of the coefficients of the vacuum polarization function 
and  $\beta$
function. 
This expansion turns out to be a critical step for the Borel transform of
the effective charge.
Introduce $\mathbf{a}^{\left(N\right)}(k^{2})$
\be
\mathbf{a}^{\left(N\right)}(k^{2}) =   \frac{1}{\frac{1}{A(k^{2})} + 
c_{1} \ln ( \mathbf{a}^{\left(N\right)}(k^{2}) ) + \sum_{2}^{N} c_{i} \mathbf{a}^{\left(N\right)}(k^{2})^{i-1}}.
\label{46} \ee
Solving $ \mathbf{a}^{\left(N\right)}(k^{2})$ in eq. (\ref{46}) 
recursively in $\alpha(\mu)$, it can be seen that this equation
generates all the higher order renormalon diagrams  to order $N$.

Let us now consider a modified Borel transform of 
$\mathbf{a}^{\left(N\right)}(
k^{2})$ defined by
\be
\mathbf{a}^{\left(N\right)}( \alpha(\mu),t)=\int_{0}^{\infty} 
\exp\left(-\frac{b}{A(t)}\right) 
\widetilde{\mathbf{a}}^{\left(N\right)}(b) \,d\,b.
\label{48} \ee
Note that this  Borel transform of GL effective charge 
was introduced by Grunberg \cite{grunberg1}.
Substituting eq. (\ref{48}) into eq. (\ref{5}), we can 
write  $D^{\left(N\right)}(\alpha(\mu),\mu^{2}/Q^{2})$,
 which is defined by replacing $\mathbf{a}(k^{2})$ in eq. (\ref{5})
 with $\mathbf{a}^{\left(N\right)}(k^{2})$, as 
\be
D^{\left(N\right)}(\alpha(\mu),\mu^{2}/Q^{2}) =  \int \exp\left[ b
 \beta_{0} \int^{\alpha(\mu)}\frac{d\, \alpha}{\beta(\alpha)}\right]
\left\{ e^{-b p_{1}} \widetilde{f}(b) \widetilde{\mathbf{a}}^{\left(N\right)}
(b)\right\} d\,b,
\label{49} \ee
where 
\be
\widetilde{f}(b) = \int_{0} f(t) \exp\left( b \beta_{0} \ln t\right) \,d\,t.
\label{50} \ee
The first IR   renormalon singularity
  arises from the IR  divergence in the integral  in eq. (\ref{50}).
 Substituting eq. (\ref{6}) into eq. (\ref{50})
\bear
\widetilde{f}(b) &=& \int_{0}^{M} f (t) e^{ b \beta_{0} \ln t} \,d\,t
 \nonumber\\
&=&-\frac{ e_{f}^{2} \mu^{4}}{8 \pi^{3} Q^{4}} \int_{0} ^{M} t 
 e^{ b \beta_{0} \ln t}
\,d\,t \nonumber \\
&=&-\frac{ e_{f}^{2} \mu^{4}}{8 \pi^{3} Q^{4}}\frac{1}{2 + b
 \beta_{0}} \left(
1 + ( 2 + b \beta_{0}) \ln M + \cdots  \right),
\label{51} \eear
where $M$  is an arbitrary
 UV  cutoff.
Notice that the leading renormalon singularity is cutoff 
independent.

The scheme dependence of  the Borel transform of $D^{\left(N\right)}(
\alpha(\mu),\mu^{2}/Q^{2})$ is now 
isolated in
\be
 e^{ - b p_{1}}, \hspace{.25in}\mbox{and} 
\hspace{.25in}  e^{b \beta_{0}
\int^{\alpha(\mu)} \frac{d\,\alpha}{\beta(\alpha)}}.
\label{53} \ee
To find the Borel transform explicitly, let us take a 
renormalization scheme in which the $\beta$ function is given by
the simple form 
\be
\beta(\alpha)=\frac{\beta_{0}  \alpha^{2}}{1-\lambda \alpha},
\label{54} \ee
where
\be
\lambda=\frac{\beta_{1}}{\beta_{0}}.
\label{55} \ee
With this $\beta$ function, eq. (\ref{49})  defines the modified 
Borel transform  by Brown, Yaffe and Zhai~(BYZ) \cite{zhai},
\be
D^{\left(N\right)}(\alpha,\mu^{2}/Q^{2})=\int_{0}^{\infty}
 \exp\left[ -b \left(\frac{1}{\alpha} +
\lambda \ln \left(\frac{\alpha}{b}\right)\right)\right] \widetilde{f}(b)
 e^{- b p_{1}} e^{-\lambda b \ln b}
\widetilde{\mathbf{a}}^{\left(N\right)}(b)
\,d\,b.
\label{56} \ee
From this we can read off the modified Borel transform of
 $D^{\left(N\right)}(\alpha(\mu),\mu^{2}/Q^{2})$
\be
\widetilde{D}^{\left(N\right)}_{\mbox{\scriptsize BYZ}}(b)= e^{- b p_{1}}\widetilde{f}(b)  
 e^{-\lambda b \ln b}
\widetilde{\mathbf{a}}^{\left(N\right)}(b).
\label{57} \ee
Then using  the relation between the ordinary Borel transform 
 and  the modified one \cite{zhai}, we have the ordinary Borel 
transform
\be
\widetilde{D}^{\left(N\right)}(b) =
-\frac{ e_{f}^{2} \mu^{4}}{16 \pi^{3} Q^{4}} e^{- b_{0} p_{1}} 
 e^{-\lambda b_{0} \ln b_{0}} \widetilde{\mathbf{a}
}^{\left(N\right)}(b_{0})
 \frac{(-2\lambda /\beta_{0})!}{(1 +\frac{1}{2} b 
\beta_{0})^{1-2\lambda/\beta_{0}}}
\left( 1 + O\left( 2  +b \beta_{0}\right)\right).
\label{58} \ee

The real problem here is to find the 
 $\widetilde{\mathbf{a}}^{\left(N\right)}(b)$.
However, with the help of the expansion in eq. (\ref{46}) it is now easy to
calculate the Borel transform of $\mathbf{a}^{\left(N\right)}(k^{2})$.
The inverse of eq. (\ref{48}) is
\be
\widetilde{\mathbf{a}}^{\left(N\right)} (b)= \frac{1}{2\pi i}\int_{ {\cal C}}
 e^{ b x } \mathbf{a}^{\left(N\right)}(x)\, d\,x,
\label{60} \ee
with the contour wrapping around the negative real axis. Here
\be
x=\frac{1}{A(k^{2})}
\label{61} \ee
and
\be
\mathbf{a}^{\left(N\right)}(x)= \frac{1}{x + c_{1} \ln \mathbf{a}^{\left(N\right)}(x) + \sum_{2}^{N}c_{i} [\mathbf{a}^{\left(N\right)}(x)]^{i-1}}.
\label{62} \ee
Putting 
\be
y=\frac{1}{\mathbf{a}^{\left(N\right)}(x)},
\label{63} \ee
we can write eq. (\ref{62}) as
\be
x=y + c_{1} \ln y - \sum_{i=2}^{N}\frac{c_{i}}{y^{i-1}}, 
\label{64} \ee
and eq. (\ref{60}) as
\be
\widetilde{\mathbf{a}}^{\left(N\right)}(b)= \frac{1}{2 \pi i} 
\int e^{b y} 
y^{ b c_{1} -1} e^{
-b \sum_{i=2}^{N}\frac{c_{i}}{y^{i-1}}} \sum_{i=0}^{N} 
\frac{\bar{c}_{i}}
{y^{i}} d\, y,
\label{66} \ee
where
\bear
\bar{c}_{i} =\left\{ \begin{array}{cc}  1 & \mbox{for $ i=0$} \\
c_{1} & \mbox{for $i=1$} \\
(i-1)\, c_{i} & \mbox{ for $ i \ge 2$}.
\end{array}
\right.
\label{67} \eear
The exponential term in the integrand may be expanded as
\be
 e^{-b \sum_{2}^{N}\frac{c_{i}}{y^{i-1}}} = \sum_{k=0}^{\infty}
\,\frac{(-b)^{k}}{k!}\, \sum_{l=k}^{k( N-1)} \frac{ h_{Nkl}}{y^{l}}
\label{68} \ee
where
\be
h_{Nkl}= k! \sum_{\left\{n_{i}\right\}} \frac{
\prod_{i=1}^{N-1} c_{i+1}^{n_{i}}}{
\prod_{i=1}^{N-1} n_{i}!}
\label{69} \ee
with the set $\{n_{i}\}$ of nonnegative integers  satisfying
\be
\sum_{i=1}^{N-1}  n_{i} \cdot i  = l, \hspace{.5in} 
\sum_{i=1}^{N-1} n_{i}= k.
\label{70} \ee
Substituting eq. (\ref{68}) into eq. (\ref{66}), we finally have
\bear
b^{bc_{1}}\widetilde{\mathbf{a}}^{\left(N\right)}(b)&=& \frac{b^{b c_{1}}}
{2 \pi i} \int e^{ b y}\, \sum_{k=0}^{\infty}
\frac{(-b)^{k}}{k!}\, \sum_{l=k}^{k (N-1)} \frac{h_{Nkl}}{y^{l}}\, \sum
_{i=0}^{N} \bar{c}_{i} y^{ b c_{1} - i-1} \,d\,y \nonumber \\
&=& \sum_{k=0}^{\infty} \,\sum_{l=k}^{k (N-1)} \,\sum_{i=0}^{N} \frac{
(-1)^{k}  h_{Nkl} \bar{c}_{i}}{
k!\, \Gamma(l+i+1-b c_{1})} b^{ k + l+i }.
\label{72} \eear
This  completes the Borel transform of the 
GL effective charge.

\section{ Renormalon residue}

To find the leading  renormalon residue of $\widetilde{D}^{\left(N\right)}(b)$,
we have to evaluate $b^{ b c_{1}}\widetilde{\mathbf{a}}^{\left(N\right)}(b)$ at
the first IR renormalon position, $b_{0}=-2/\beta_{0}$. If we directly
substitute $b$ in eq. (\ref{72}) with $b_{0}$, the resulting large order behavior
would sum all the contribution from the renormalons to order $N$, but
 unfortunately
this large order behavior does not have a  finite limit for
 $N \rightarrow \infty$ \cite{fs}. The reason for this is that  $
\widetilde{\mathbf{a}}(b) $ is singular at the  UV and IR renormalon
positions, and its radius of convergence when it is expanded as in
 eq. (\ref{72})
is given by the position at $b=1/\beta_{0}$  of the first UV renormalon, 
which is the closest renormalon to the origin in  the Borel plane.
 Therefore we cannot
 substitute $b$ with $b_{0}$ in eq. (\ref{72}) 
to correctly  evaluate the 
Borel transform  at the first
IR renormalon.

This problem can be avoided by introducing an analytic transform of the
Borel plane so that the closest renormalon to the origin in the
new complex plane  is the first IR renormalon \cite{mueller3}.
Because the singularity of  $\widetilde{\mathbf{a}}(b) $ at the IR 
renormalon is
such that it is finite but has divergent derivative \cite{grunberg1},
we can then express the residue as a convergent series.

For this purpose,
 we can take any analytic transform that puts the IR renormalon
as the closest singularity to the origin, but here 
we consider a simple form
\be
z= \frac{\beta_{0} b}{1-\beta_{0} b},
\label{sun2}\ee
with its inverse
\be
b=\frac{1}{\beta_{0}} \left( \frac{z}{1+z}\right).
\label{s1}
\ee
In the $z-\mbox{plane}$, the closest singularity to the origin is the
first IR renormalon at
\be
z_{0}=-\frac{2}{3},
\ee
and all the UV renormalons are pushed beyond $z=-1$ on the real axis.
It is interesting to note that the  freedom in choosing 
this analytic transform is similar to
that for the  renormalization scheme. As the renormalization scheme 
can be optimized for a particular process, 
 the  analytic transform could be
chosen to optimize the perturbative evaluation of the residue.

Now to find  $ b^{bc_{1}}
\widetilde{\mathbf{a}}^{\left(N\right)}(b)$ at the first IR renormalon,
we have to substitute $b$ in eq. (\ref{72}) with that in 
 eq. (\ref{s1}) and expand it  in
Taylor series at $z=0$ to order $N$, and  evaluate it at
$z=z_{0}$.
Thus the Borel transform  of GL effective charge at the first IR renormalon
is given by
\bear
 \kappa_{N}&=&\left. b^{bc_{1}} \widetilde{\mathbf{a}}^{\left(N\right)}(b)
 \right|_{b=b_{0}} \nonumber \\
&=&\sum_{M=0}^{N} q_{M}\, z_{0}^{M} ,
\label{10001}
\eear
with
\be
q_{M}= (M-1)!\beta_{0}^{-M} \sum_{k=0}^{N} 
\sum_{l=k}^{k (N-1)} \sum_{i=0}^{N}
\sum_{j=0}^{N}\sum_{m=0}^{N}  \frac{
(-1)^{k+m}  h_{Nkl} \bar{c}_{i} c_{1}^{j}}{ k! m! (M-m-1)!}\gamma_{j}^{
\left\{l+i\right\}} \beta_{0}^{m} \delta_{k+l+i+j+m,M},
\ee
Here $\gamma_{j}^{\left\{n\right\}}$ is defined by 
\be
\frac{1}{\Gamma(n+1-x)}=\sum_{j=0} \gamma_{j}^{\left\{n\right\}} x^{j}.
\ee
Note that the coefficient $q_{N}$ is completely determined by
$\widetilde{\mathbf{a}}^{\left(N\right)}(b)$ in eq. (\ref{72}), and is not
modified by the  Borel transform 
$\widetilde{\mathbf{a}}^{\left(N+m\right)}(b)$ of the higher order effective 
charge. The IR residue is then
\be
 \kappa =\sum_{M=0}^{\infty} q_{M}\, z_{0}^{M}.
\ee
It should be emphasized that, though this series  is being evaluated at
 its radius of convergence, it is convergent 
because of the finiteness of 
 $\widetilde{\mathbf{a}}(b(z))$  at $z=z_{0}$.
 
\section{ Large order behavior}

The Borel transform  $\widetilde{D}(b)$  can now  be
used in determining the leading large order behavior of $D(\alpha)$.
 Note that
the leading large order behavior is determined by the
leading  Borel singularity,  and  the $1/n$ correction in the
large order behavior corresponds to the $O( b-b_{0})$ correction in
the Borel transform. First,
we give  the large order behavior in a renormalization
scheme in which the $\beta$ function is given by eq. (\ref{54}),
and then  in  then sec. 7  discuss a class of schemes in which
all schemes share a common large order behavior except for a 
trivial scheme dependent term.
Using the known result for the vacuum polarization function
and the $\beta$ function to three-loop and four-loop respectively, we
will also give numerical values for the large order behavior.

Let us go back to the Borel transform of $D$ in eq. (\ref{58}).
Expanding $\widetilde{D}^{\left(N\right)}(b)$ at  the origin, and using the
definition of the Borel transform in eq. (\ref{36}), the leading large order
 behavior of $D^{\left(N\right)}(\alpha(\mu),\mu^{2}/Q^{2})$  is
 given by
\be
a^{\left(N\right)}_{n} = -\frac{ e_{f}^{2} \mu^{4}}{16 \pi^{3} Q^{4}}
 e^{- b_{0}p_{1} } e^{-\lambda b_{0}
 \ln b_{0}} \widetilde{\mathbf{a}}^{\left(N\right)}\left(b_{0}\right)
 n! n^{\lambda b_{0}} b_{0}^{-n} \hspace{.25in}\mbox{ with $b_{0}=
\frac{-2}{\beta_{0}}$}. 
\label{73} \ee
Then the sequence for the large order behavior mentioned in sec. 1
may be defined as 
\be
K_{N}  = -\frac{e_{f}^{2} \mu^{4}}{16 \pi^{3} Q^{4}} e^{
 - b_{0} p_{1}}  \kappa_{N}
\label{76} \ee
with $\kappa_{N}$ defined in eq (\ref{10001}).

To evaluate the numerical values for $\kappa_{N}$, we have to find
the coefficients $c_{i}$ in eq. (\ref{46}) explicitly. The 
$\beta$ function and the vacuum polarization function
 in $\overline{\mbox{MS}}$ scheme to three-loop 
and four-loop respectively are
given by \cite{kataev}
\bear
\beta_{0} &=& \frac{1}{2 \pi} \left(\frac{2}{3} N_{f}\right) \nonumber \\
\beta_{1} &=& \frac{1}{2 \pi^{2}} \left(\frac{N_{f}}{2} \right) \nonumber \\
\beta_{2} &=& \frac{1}{2 \pi^{3}} \left(-\frac{1}{16} N_{f} -
\frac{11}{72} N_{f}^{2}\right) \nonumber \\
\beta_{3} &=& \frac{1}{2 \pi^{4}} \left[-\frac{23}{64} N_{f} +
\left(\frac{95}{432} -\frac{13}{18} \zeta(3)\right) N_{f}^{2} -
\frac{77}{1944} N_{f}^{3}\right] 
\label{78} \eear
and
\bear
p_{1}& =& \frac{5}{9 \pi} N_{f} \nonumber \\
p_{2} &=& \frac{N_{f}}{\pi^{2}}\left[ \frac{55}{48} -\zeta(3)\right] 
 \nonumber \\
p_{3} &=&\frac{1}{\pi^{3}} \left[ \left( -\frac{143}{288} 
-\frac{37}{24}\zeta(3) +\frac{5}{2}\zeta(5)\right) N_{f} +
 \left(-\frac{3701}{2592} + \frac{19}{18} 
\zeta(3)\right)  N_{f}^{2} \right],
\label{79} \eear 
where  $N_{f}$ is the number of fermion flavors.
Solving $\mathbf{a}(k^{2})$ in eq. (\ref{38}) in terms of $\alpha(\mu)$ and comparing it with eq. (\ref{16})
we find
\bear
c_{1}&=&- \frac{\beta_{1}}{\beta_{0}}\nonumber \\
c_{2}&=&-\frac{\beta_{2}}{\beta_{0}} +\frac{\beta_{1}^{2}}{\beta_{0}^{2}} -
\frac{p_{1} \beta_{1}}{\beta_{0}} + p_{2} \nonumber \\
c_{3}&=&-\frac{\beta_{3}}{2 \beta_{0}} +\frac{\beta_{1}\beta_{2}}
{\beta_{0}^{2}} -
\frac{\beta_{1}^{3} }{2 \beta_{0}^{3}}  -
 \frac{p_{1}\beta_{2}}{\beta_{0}} 
+ \frac{p_{1} \beta_{1}^{2}}{\beta_{0}^{2}}  \nonumber \\
& &  - \frac{p_{1}^{2} \beta_{1}}{2 \beta_{0}} - \frac{p_{2} \beta_{1}}{\beta_{0}} + p_{1} p_{2} +p_{3}
\label{81} \eear
Note that the use of the vacuum polarization function and the
$\beta$ function in $\overline{\mbox{MS}}$ is allowed, because
$c_{i}$ are independent of renormalization scheme. 

The $\kappa_{N}$ obtained by substituting eq. (\ref{81})
 into eq. (\ref{10001}) is given in Table 1 for several  flavor numbers.
In the table, we see that for a  reliable estimation of the
large order behavior,  a higher order calculation beyond the current one
is required for the vacuum polarization function and the 
$\beta$ function.

Not surprisingly, the 
 numbers in the table also suggest that the large $N_{f}$ limit
is the one-loop renormalon. This is indeed the case. To see this,
note that the following coefficients scale as
\bear
\beta_{i} &\sim& N_{f}^{i} \hspace{.5in} \mbox{for\,\, $ i >0$}, 
\nonumber \\ 
c_{i} &\sim& N_{f}^{i-1}
\eear
for large $N_{f}$.
Then scaling the variable $y$ in eq. (\ref{66}) by $b_{0}\, y$, it is 
straightforward to see that 
\be
\lim_{N_{f}\rightarrow \infty} \kappa_{N} = 1.
\ee 
\begin{table}
\centering
\begin{tabular}{|c|c|c|c|c|c|c|} \hline
& $N_{f}=1$ & $N_{f}=2$ & $N_{f}=3$ & $N_{f}=4$ & $N_{f}=5$ &$N_{f}=100$ \\ \hline
$\kappa_{0}$ &1  &1 &1 &1 &1&1 \\ \hline
$\kappa_{1}$ &1.63  &1.32 &1.21 &1.16 &1.13&1.00 \\ \hline
$\kappa_{2}$ &0.71  &1.31 &1.31 &1.27 &1.24&1.02 \\ \hline
$\kappa_{3}$ &-1.53 &1.25 &1.41 &1.39 &1.34&1.02 \\ \hline
\end{tabular}
\isucaption{
 The first three elements of the sequence for the 
first IR renormalon residue in QED. 
 $\kappa_{0}$ denotes the residue from the one-loop
renormalon.}
\end{table} 

\section{ Residue in QCD}

In QCD,  there is unfortunately 
no satisfactory definition of renormalization 
scheme and scale invariant effective charge that may be used in the
diagrammatic study of renormalon. 
However, as long as such an effective charge is defined, the
formalism developed in QED may be used without modification.

Often in renormalon calculation in QCD, $\Pi(t)$ in eq. (\ref{16}) that 
defines the effective charge is considered in certain limit\cite{sirlin},
 for example, as  in the $1/N_{f}$ approximation
combined with  ``naive-nonabelianization'' \cite{NA}, and
the pinch technique  \cite{PT}.
The pinch technique appears to be promising, though 
presently there is no all-order definition for the effective charge in this
scheme. In pinch technique,
$\Pi(t)$ at one-loop level is defined by collecting the gluon
vacuum polarization, and the vacuum polarization-like term in the
vertex and box diagrams.

However, if we are only interested in the calculation of the 
residue, the definition
of the effective charge is not required. Indeed the calculation 
 is cunningly simple; it only requires the strength of the 
renormalon singularity and the perturbative calculation of $D(\alpha)$.

Consider the Borel transform of the current correlation function in QCD.
The renormalon singularity of   $\widetilde{D}(b)$ in QCD
\be
 \widetilde{D}(b) \approx \frac{\widehat{D}}{ (1- b/b_{0})^{ 1 + 
\lambda b_{0}}}
\label{t1}
\ee
gives the large order behavior
\be
a_{n} \approx \frac{\widehat{D}}{(\lambda\, b_{0} )!} n! n^{\lambda\, b_{0}}
\, b_{0}^{-n}.
\ee

To calculate the residue $\widehat{D}$, consider a function
\be 
R(b) = \widetilde{D}(b)\, (1-b/b_{0})^{ 1 + \lambda b_{0}}.
\ee
Then because of eq. (\ref{t1}), we have
\be
\widehat{D} = R(b_{0}).
\ee
To avoid the first UV renormalon,  we introduce  a new variable $z$, as we
did in QED, 
which is defined by
\be
z= -\frac{\beta_{0} b}{1-\beta_{0} b}
\ee
with its inverse
\be
b= \frac{-1}{\beta_{0}} \left( \frac{z}{1-z}\right).
\ee
In the $z-$plane, the IR renormalon at
\be 
z_{0}=\frac{2}{3}
\ee
is the closest singularity to the origin, and so the radius of convergence
of the Taylor series of  $\widetilde{D}(b(z))$ at $z=0$
is given by the first IR renormalon. 

Now $ \widehat{D}$
can be expressed in a convergent series form
\bear
\widehat{D} &=&\left. \widetilde{D}(b)\, (1-b/b_{0})^{ 1 + 
\lambda b_{0}}\right|_{b=b_{0}} \nonumber \\
&=& \left.\left(\sum_{n=0}^{\infty} \frac{a_{n}}{n!} \left[b(z)\right]^{n}
\right) (1-b(z)/b_{0})^{ 1 + \lambda b_{0}}\right|_{z=z_{0}} \nonumber \\
&=& \sum_{n=0}^{\infty} r_{n} z_{0}^{n},
\eear
where it is straightforward to find $r_{n}$ in terms of the 
perturbative coefficients $a_{n}$. Note that the series is 
convergent even if $R(b(z))$ is not analytic at $z=z_{0}$, because then
the radius of convergence of the series is given by $z=z_{0}$, and 
$R(b(z_{0}))$ is finite.

Using the  perturbative calculation of the current correlation
function, and $D(\alpha)$, to three loop \cite{zhai}, we have
\be
R(b(z))= \frac{3 \sum_{f} Q_{f}^{2}}{ 16\, \pi^3} \left[
1.333-0.748\,z -0.311 \,z^{2} +O(z^{3})\right]
\ee
for $N_{f}=3$. This is in the renormalization scheme in which the 
one-loop renormalization point is same as that of $\overline{MS}$ scheme,
and  the $\beta$
function is given in the form in eqs. (\ref{54}), (\ref{55}).
Evaluating this series at the renormalon position at $z=z_{0}$
we have
\bear
K_{1}&=&\frac{1.333}{(\lambda b_{0})!}= 0.946 \nonumber \\
K_{2}&=&\frac{(1.323-0.748\,z_{0})}{(\lambda b_{0})!}=0.592  \nonumber \\
K_{3}&=&\frac{(1.323-0.748\,z_{0}-0.311 \,z_{0}^{2})}
{(\lambda b_{0})!}= 0.494  \nonumber \\
\eear
For  several other flavor numbers we give $K_{n}$   in Table 2.
\begin{table}
\centering
\begin{tabular}{|c|c|c|c|c|c|} \hline
& $N_{f}=1$ & $N_{f}=2$ & $N_{f}=3$ & $N_{f}=4$ & $N_{f}=5$ \\ \hline
$K_{1}$ &.881  &.904 &.946 &1.018 &1.132  \\ \hline
$K_{2}$ &.521  &.546 &.592 &.674 &.813  \\ \hline
$K_{3}$ &.592 & .549 &.494&.411 &.307 \\ \hline
\end{tabular}
\isucaption{
 The first three elements of the sequence for the 
large order behavior in QCD.}
\end{table} 

\section{ Scheme dependence of large-order behavior}

In section 5, we determined the large order behavior in 
renormalization schemes in which the $\beta$ function is given by
the simple form  in eq. (\ref{54}). With this $\beta$ function, the scheme
dependence of the  large order behavior arises only through the factor
\be
e^{- b_{0} p_{1}}
\label{82} \ee
in eq. (\ref{76}).
 In fact, this result is more  general. All renormalization 
schemes for which the coefficients of the  $\beta$ function do not 
grow faster than $ a_{n}$
share the same large-large order behavior except for the scheme 
dependence in eq. (\ref{82}).

To see this, let us consider two renormalization schemes (say,
unprimed and primed ) in which
the relation between the couplings is given by eq. (\ref{43}).
We now assume that the large order behavior by the first UV renormalon
 is extracted out so that the leading large order behavior 
of a scheme independent subamplitude $D'$ of $D$
  is given by the first IR renormalon.
 Then using
\be
D' = \sum_{n=0}^{\infty} a_{n} \alpha(\mu)^{n+1} =
\sum_{n=0}^{\infty}a'_{n}  \alpha'(\mu)^{n+1},
\label{83} \ee
we get the relation between $a_{n}$ and $a'_{n}$
\be
a_{n} = a'_{n} \left( 1 + (n+1) \frac{a'_{n-1}}{a'_{n}} \gamma_{1}
+\frac{n(n+1)}{2} \frac{ a'_{n-2}}{a'_{n}} \gamma_{1}^{2} +
\cdots + \frac{a'_{0} \gamma_{n}}{a'_{n}} \right).
\label{84} \ee
With the large order behavior of $a'_{n}$ in the form in eq. (\ref{1}),
 this equation becomes 
\be
a'_{n}=a_{n} \, e^{- b_{0} \gamma_{1}} \left( 1 +
O\left(1/n\right)\right) 
\label{85} \ee
provided
\be
\lim_{n \rightarrow \infty} \frac{n \gamma_{n}}{a'_{n}} = \mbox{const.}
\label{86} \ee
Thus if $\gamma_{n}$ does not grow faster than
\be
(n-1)! n^{\nu} b_{0}^{-n},
\label{87} \ee
the scheme dependence of large order behavior is given by the
simple relation in eq. (\ref{85}). In fact the large order behavior we 
found in  sec. 5 exactly transforms according to eq. (\ref{85})
under scheme changes. 

We may now translate the limit in eq. (\ref{86}) on $\gamma_{n}$ to that
 of $\beta_{n}$.
Let the $\beta$ function in the unprimed scheme is given by
the simple form in eq. (\ref{54}). Then eq. (\ref{43}) gives
  the $\beta'(\alpha')$ in the
form
\bear
\beta'(\alpha') =\sum_{n=0}^{\infty} \beta'_{n} \alpha'^{n+2} 
&=&\beta(\alpha) \sum_{n=0}^{\infty} (n+1) \gamma_{n} \alpha^{n} \nonumber \\
&=& \frac{\beta_{0} \alpha^{2}}{ 1- \lambda \alpha} \sum_{n} ( n+1) \gamma_{n} \alpha^{n} \nonumber \\
&=& \beta_{0} \sum_{n=0}^{\infty} \bar{\beta'_{n}}  \alpha^{n+2}
\label{88} \eear
where
\bear
 \bar{\beta'_{n}}&=& \sum_{k=0}^{n} \lambda^{n-k}\,(k+1)\,
\gamma_{k} \nonumber \\
&\sim& n \,\gamma_{n}\,( 1+ O(1/n))
\label{89} \eear
Inverting eq. (\ref{43}) to express $\alpha$ in terms of $\alpha'$, 
and substituting
it in eq. (\ref{88}), we have 
\be
\beta'_{n} =\bar{\beta'}_{n}( 1+O(1/n))=n \gamma_{n} (1 + O(1/n)).
\label{90} \ee
Then the restriction on $\gamma_{n}$ in eq. (\ref{87}) implies that
any renormalization scheme in which $\beta_{n}$ does
not grow faster than
\be
n! \,n^{\nu} \,b_{0}^{-n}
\label{91} \ee
has the same large order behavior (except for 
the factor eq. (\ref{82}))
 as in a scheme for which
the $\beta$ function is given by eq. (\ref{54}).

\section{Acknowledgements}
\noindent
This work arose from working with A. Mueller several years ago.
I am deeply grateful to him for discussions from which this work
was motivated. I am also greatly indebted to Chengxing Zhai for
stimulating discussions and wish to thank T. Clark and S. Love for
valuable comments. After completion of the paper I came across with the work by S. Faleev and
P. Silvestrov \cite{fs} in which part of our subject is  also discussed.
I thank A. Mueller for bringing this paper to my attention.


\end{document}